\documentclass[runningheads]{llncs}

\usepackage{multirow}
\usepackage{marvosym}

\usepackage{graphicx}
\usepackage[caption=false]{subfig}

\usepackage{array}
\newcommand{\PreserveBackslash}[1]{\let\temp=\\#1\let\\=\temp}
\newcolumntype{C}[1]{>{\PreserveBackslash\centering}p{#1}}

\begin{document}              

\title{What Drives Readership? An Online Study on User Interface Types and Popularity Bias Mitigation in News Article Recommendations}
\titlerunning{Interface Types and Popularity Bias in News Article Recommendations}  
%


\author{Emanuel Lacic \inst{1} \and Leon Fadljevic \inst{1} \and Franz Weissenboeck \inst{2} \and Stefanie Lindstaedt \inst{1} \and Dominik Kowald \inst{1}$^{\textrm{(\Letter)}}$}
\authorrunning{E. Lacic, L. Fadljevic, F. Weissenboeck, S. Lindstaedt, and D. Kowald}

\institute{Know-Center GmbH, Graz, Austria \\\email{{elacic,lfadljevic,slind,dkowald}@know-center.at} \and
Die Presse Verlags-GmbH, Vienna, Austria \\\email{franz.weissenboeck@diepresse.com}}

\maketitle


\begin{abstract}
Personalized news recommender systems support readers in finding the right and relevant articles in online news platforms. 
In this paper, we discuss the introduction of personalized, content-based news recommendations on DiePresse, a popular Austrian online news platform, focusing on two specific aspects: (i) user interface type, and (ii) popularity bias mitigation.
Therefore, we conducted a two-weeks online study that started in October 2020, in which we analyzed the impact of recommendations on two user groups, i.e., anonymous and subscribed users, and three user interface types, i.e., on a desktop, mobile and tablet device. With respect to user interface types, we find that the probability of a recommendation to be seen is the highest for desktop devices, while the probability of interacting with recommendations is the highest for mobile devices. With respect to popularity bias mitigation, we find that personalized, content-based news recommendations can lead to a more balanced distribution of news articles' readership popularity in the case of anonymous users. Apart from that, we find that significant events (e.g., the COVID-19 lockdown announcement in Austria and the Vienna terror attack) influence the general consumption behavior of popular articles for both, anonymous and subscribed users.


\end{abstract}




\section{Introduction}

Similar to domains such as social networks or social tagging systems~\cite{lacic2014recommending,seitlinger2015attention,kowald2013social}, the personalization of online content has become one of the key drivers for news portals to increase user engagement and convince readers to become paying subscribers~\cite{garcin2013pen,garcin2014swissinfo,deSouza2018DNN}. 
A natural way for news portals to do this, is to provide their users with articles that are fresh and popular. This is typically achieved via simple most-popular news recommendations, especially since this approach has been shown to provide accurate recommendations in offline evaluation settings~\cite{kille2015overview}. However, such an approach could amplify popularity bias with respect to users' news consumption. This means that the equal representation of non-popular, but informative content in the recommendation lists is put into question, since articles from the ``long tail'' do not have the same chance of being represented and served to the user~\cite{abdollahpouri2019popularity}. 
Since nowadays, readers tend to consume news content on smaller user interface types (e.g., mobile devices)~\cite{KARIMI20181203,newman2015reuters}, the impact of popularity bias may even get amplified due to the reduced number of recommendations that can be shown~\cite{kim2015eye}.

In this paper, we therefore discuss the introduction of personalized, content-based news articles on DiePresse, a popular Austrian news platform, focusing on two aspects: (i) user interface type, and (ii) popularity bias mitigation. 
To do so, we performed a two-weeks online study that started in October 2020, in which we compared the impact of recommendations with respect to different user groups, i.e., anonymous (cold-start~\cite{lacic2015tackling}) and subscribed (logged-in and paying) users, as well as different user interface types, i.e., desktop, mobile and tablet devices (see Section~\ref{sec:exp_setup}). Specifically, we address two research questions:

\vspace{1mm}
\noindent
\textbf{RQ1:} How does the user interface type impact the performance of news recommendations? 

\noindent
\textbf{RQ2:} Can we mitigate popularity bias by introducing personalized, content-based news recommendations?

\vspace{1mm}
\noindent
We investigate RQ1 in Section~\ref{sec:rq1} and RQ2 in  Section~\ref{sec:rq2}. Additionally, we discuss the impact of two significant events, i.e., (i) the COVID-19 lockdown announcement in Austria, and (ii) the Vienna terror attack, on the consumption behavior of users. We hope that our findings will help other news platform providers assessing the impact of introducing personalized  recommendations.

\section{Experimental Setup} \label{sec:exp_setup}


In order to answer our two research questions, we performed a two-weeks online user study, which started on the 27th of October 2020 and ended on the 9th of November 2020. Here, we focused on three user interface types, i.e., desktop, mobile and tablet devices, as well as investigated two user groups, i.e., anonymous and subscribed users. About $89\%$ of the traffic (i.e., $2,371,451$ user interactions) was produced by the 1,182,912 anonymous users, where a majority of them (i.e., $77.3\%$) read news articles on a mobile device. 
Interestingly, the 15,910 subscribed users exhibited a more focused reading behavior and only interacted with a small subset of all articles that were read during our online study (i.e., around $18.7\%$ out of $17,372$ articles). Within the two-weeks period, two significant events happened: (i) the COVID-19 lockdown announcement in Austria on the 31st of October 2020, and (ii) the Vienna terror attack on the 2nd of November 2020. The articles related to these events were the most popular ones in our study.

\vspace{1mm}
\noindent
\textbf{Calculation of Recommendations.}
We follow a content-based approach to recommend news articles to users~\cite{lops2011content}. Therefore, we represent each news article using a 25-dimensional topic vector calculated using Latent Dirichlet Allocation (LDA)~\cite{blei2003latent}. Each user was also represented by a 25-dimensional topic vector, where the user's topic weights are calculated as the mean of the news articles' topic weights read by the user. In case of subscribed users, the read articles consist of the entire user history and in case of anonymous users, the read articles consist of the articles read in the current session.
Next, these topic vectors are used to match users and news articles using Cosine similarity in order to find top-$n$ news article recommendations for a given user. For our study, we set $n = 6$ recommended articles. For this step, only news articles are taken into account that have been published within the last 48 hours. Additionally, editors had the possibility to also include older (but relevant) articles into this recommendation pool (e.g., a more general article describing COVID-19 measurements).

In total, we experimented with four variants of our content-based recommendation approach: (i) recommendations only including articles of the last 48 hours, (ii) recommendations also including the editors' choices, and (iii) and (iv) recommendations, where we also included a collaborative component by mixing the user's topic vector with the topic vectors of similar users for the variants (i) and (ii), respectively. Additionally, we also tested a most-popular approach, since this algorithm was already present in DiePresse before the user study started. However, we did not find any significant differences between these five approaches with respect to recommendation accuracy in our two-weeks study and therefore, we did not distinguish between the approaches and report the results for all calculated recommendations in the remainder of this paper.

\section{Results}
\subsection{RQ1: User Interface Type} \label{sec:rq1}

Most studies focus on improving the accuracy of the recommendation algorithms, but recent research has shown that this has only a partial effect on the final user experience~\cite{knijnenburg2012explaining}. 
The user interface is namely a key factor that impacts the usability, acceptance and selection behavior within a recommender system~\cite{felfernig2012preface}. Additionally, in news platforms, we can see a trend that shifts from classical desktop devices to mobile ones. 
Moreover, users are biased towards clicking on higher ranked results (i.e., position bias)~\cite{craswell2008experimental}. When evaluating personalized news recommendations, it becomes even more important to understand the user acceptance of recommendations for smaller user interface types, where it is much harder for the user to see all recommended options due to the limited size. 
In our study, we therefore investigate to what extent the user interface type impacts the performance of news recommendations (RQ1). As mentioned, we differentiate between three different user interface types, i.e., interacting with articles on a (i) desktop, (ii) mobile, and (iii) tablet device. In order to measure the acceptance of recommendations shown via the chosen user interface type, we use the following two evaluation metrics~\cite{garcin2014swissinfo}:

\vspace{1mm}
\noindent
\textbf{Recommendation-Seen-Ratio (RSR)} is defined as the ratio between the number of times the user actually saw recommendations (i.e., scrolled to the corresponding recommendation section in the user interface) and the number of recommendations that were generated for a user.

\noindent
\textbf{Click-Through-Rate (CTR)} is measured by the ratio between the number of actually clicked recommendations and the number of seen recommendations.

\vspace{1mm}
\noindent
As shown in Table~\ref{tab:ui_choice}, the smaller user interface size of a mobile device heavily impacts the probability of a user to actually see the list of recommended articles. This may be due to the fact that reaching the position where the recommendations are displayed is harder in comparison to a larger desktop or tablet device, where the recommendation section can be reached without scrolling. Interestingly enough, once a user has seen the list of recommended articles, users who use a mobile device exhibit a much higher CTR.
Again, we hypothesize that if a user has put more effort into reaching the list of recommended articles, the user is more likely to accept the recommendation and interact with it. 

\begin{figure}[t]
\centering

\subfloat[Recommendation-Seen-Ratio.]{
\includegraphics[width=.46\textwidth]{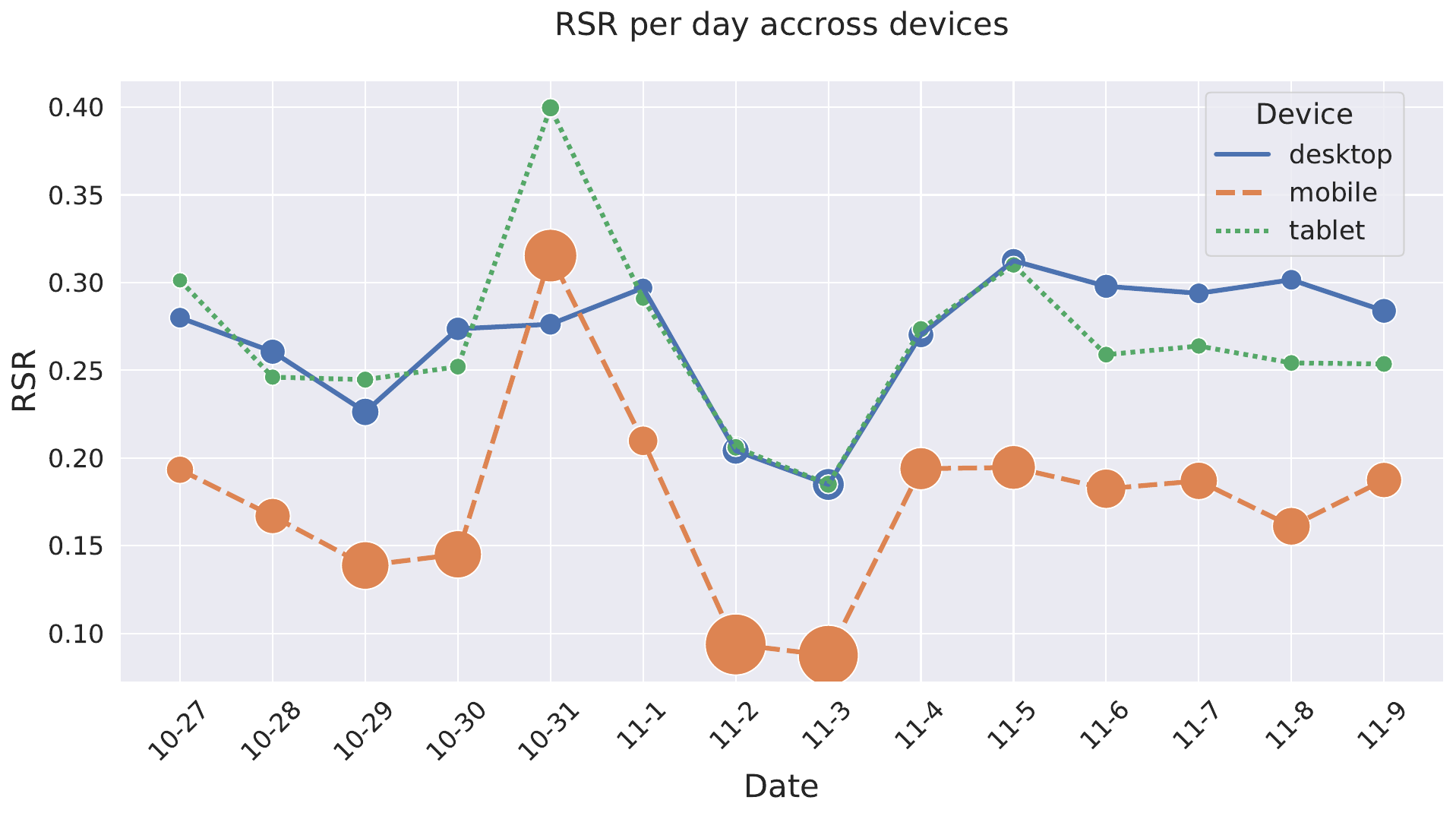}}
~
\subfloat[Click-Through-Rate.]{
\includegraphics[width=.46\textwidth]{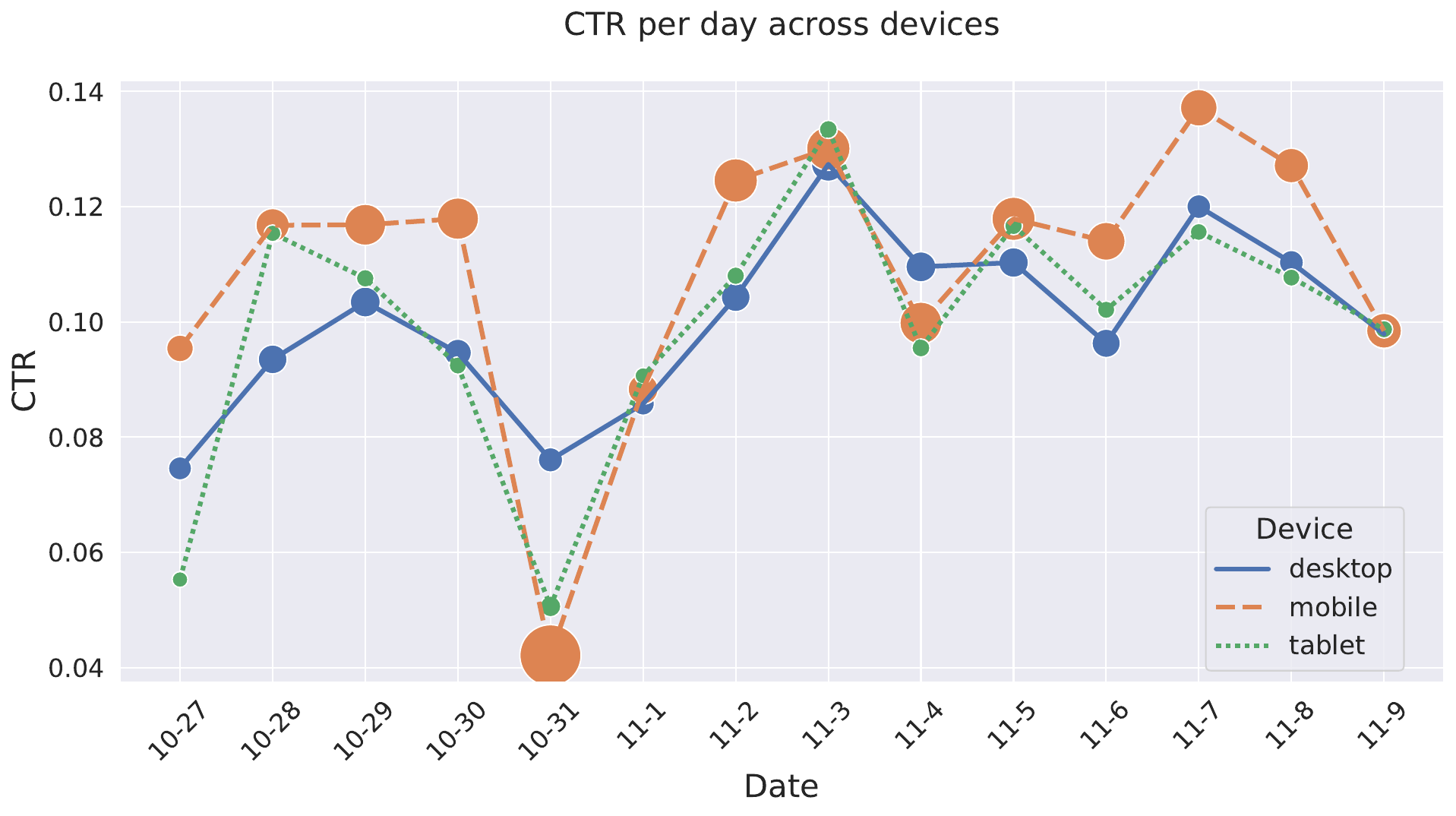}}

\caption{RQ1: Acceptance of recommended articles for the two weeks of our study with respect to (a) RSR, and (b) CTR. The size of the dots represent the number of reading events on a specific day for a specific user interface type.}
\vspace{-5mm}
\label{fig:ui}
\end{figure}

When looking at Figure~\ref{fig:ui}, we can see a consistent trend during the two weeks of our study regarding the user interface types for both the RSR and CTR measures. However, notable differences are the fluctuations of the evaluation measures for the two significant events that happened during the study period. For instance, the positive peak in the RSR and the negative peak in CTR that can be spotted around the 31st of October was caused by the COVID-19 lockdown announcement in Austria. For the smaller user interfaces (i.e., mobile and tablet devices) this actually increased the likelihood of the recommendation to be seen since users have invested more energy in engaging with the content of the news articles. On the contrary, we saw a drop in the CTR, which was mostly caused by anonymous users since the content-based, personalized recommendations did not provide articles that they expected at that moment (i.e., popular ones solely related to the event).
Another key event can be spotted on the 2nd of November, the day the Vienna terror attack happened. 
This was by far the most read article with a lot of attack-specific information during the period of the online study.
Across all three user interface types, this has caused a drop in the likelihood of a recommendation to be seen at all. Interestingly enough, the CTR in this case does not seem to be influenced. We investigated this in more detail and noticed that a smaller drop was only noticeable for the relatively small number of subscribed users using a mobile device and thus, this does not influence the results shown in Figure~\ref{fig:ui}.
The differences between all interface types shown in Table~\ref{tab:ui_choice} and Figure~\ref{fig:ui} are statistically significant according to a Kruskal-Wallis followed by a Dunn test except for mobile vs. tablet device with respect to CTR.

\def\arraystretch{1}%
\begin{table}[t!]
\caption{RQ1: Acceptance of recommended articles with respect to user interface type.\vspace{-2mm}} 
\begin{tabular}{r|C{2cm}|C{2cm}|C{2cm}}
\hline 
  Metric      & Desktop & Mobile & Tablet \\ \hline
RSR: Recommendation-Seen-Ratio (\%) & \textbf{26.88}   & 17.55  & 26.71  \\ \hline
CTR: Click-Through-Rate (\%) &    10.53     &    \textbf{13.40}    &  11.37      \\ \hline 
\end{tabular}
\vspace{-3mm}
\label{tab:ui_choice}
\end{table}
\def\arraystretch{1}%

\subsection{RQ2: Mitigating Popularity Bias}  \label{sec:rq2}

Many recommender systems are affected by popularity bias, which leads to an overrepresentation of popular items in the recommendation lists. One potential issue of this is that unpopular items (i.e., so-called long-tail items) are recommended rarely~\cite{kowald2021support,kowald2020unfairness}. 
The news article domain is an example where ignoring popularity bias could have a significant societal effect. For example, a potentially controversial news article could easily impose a narrow ideology to a large population of readers~\cite{flaxman2016filter}. This effect could even be strengthened by providing unpersonalized, most-popular news recommendations as it is currently done by many online news platforms (including DiePresse) since these popularity-based approaches are easy to implement and also provide good offline recommendation performance~\cite{garcin2014swissinfo,KARIMI20181203}. 
We hypothesize that the introduction of personalized, content-based recommendations (see Section~\ref{sec:exp_setup}) could lead to more balanced recommendation lists in contrast to most-popular recommendations. This way also long-tail news articles are recommended and thus, popularity bias could be mitigated. 
Additionally, we believe that this effect differs between different user groups and thus, we distinguish between anonymous and subscribed users.

We measure popularity bias in news article consumption by means of the skewness~\cite{bellogin2017statistical} of the article popularity distribution, i.e., the distribution of the number of reads per article. Skewness measures the asymmetry of a probability distribution, and thus a high, positive skewness value depicts a right-tailed distribution, which indicates biased news consumption with respect to article popularity. On the contrary, a small skewness value depicts a more balanced popularity distribution with respect to head and tail, and thus indicates that also non-popular articles are read. As another measure, we calculate the kurtosis of the popularity distribution, which measures the ``tailedness'' of a distribution. Again, higher values indicate a higher tendency for popularity bias.
For both metrics, we hypothesize that the values at the end of our two-weeks study are smaller than at the beginning, which would indicate that the personalized recommendations helped to mitigate popularity bias.

\begin{figure}[t]
\centering

\subfloat[Skewness.]{
\includegraphics[width=.46\textwidth]{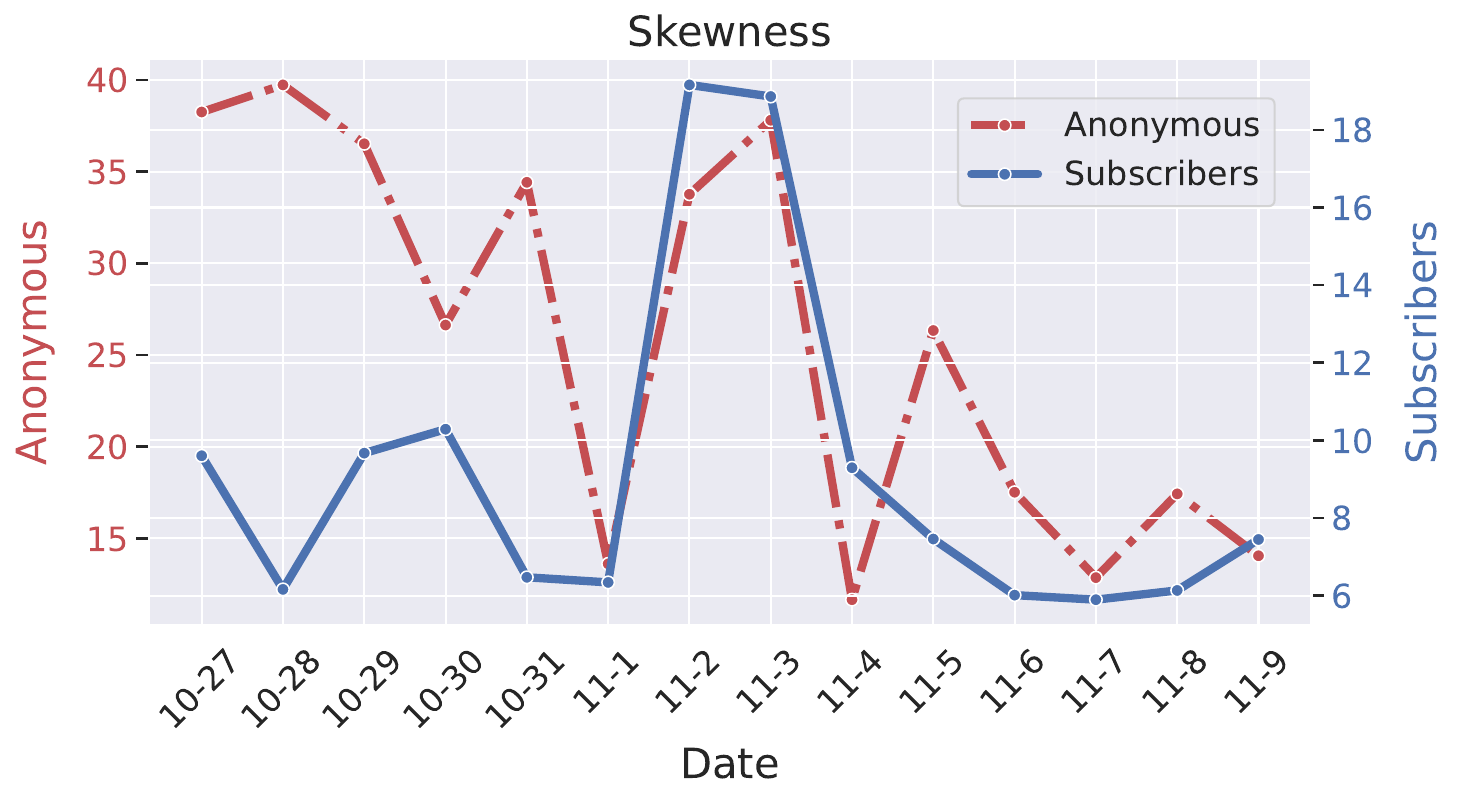}}
~
\subfloat[Kurtosis.]{
\includegraphics[width=.46\textwidth]{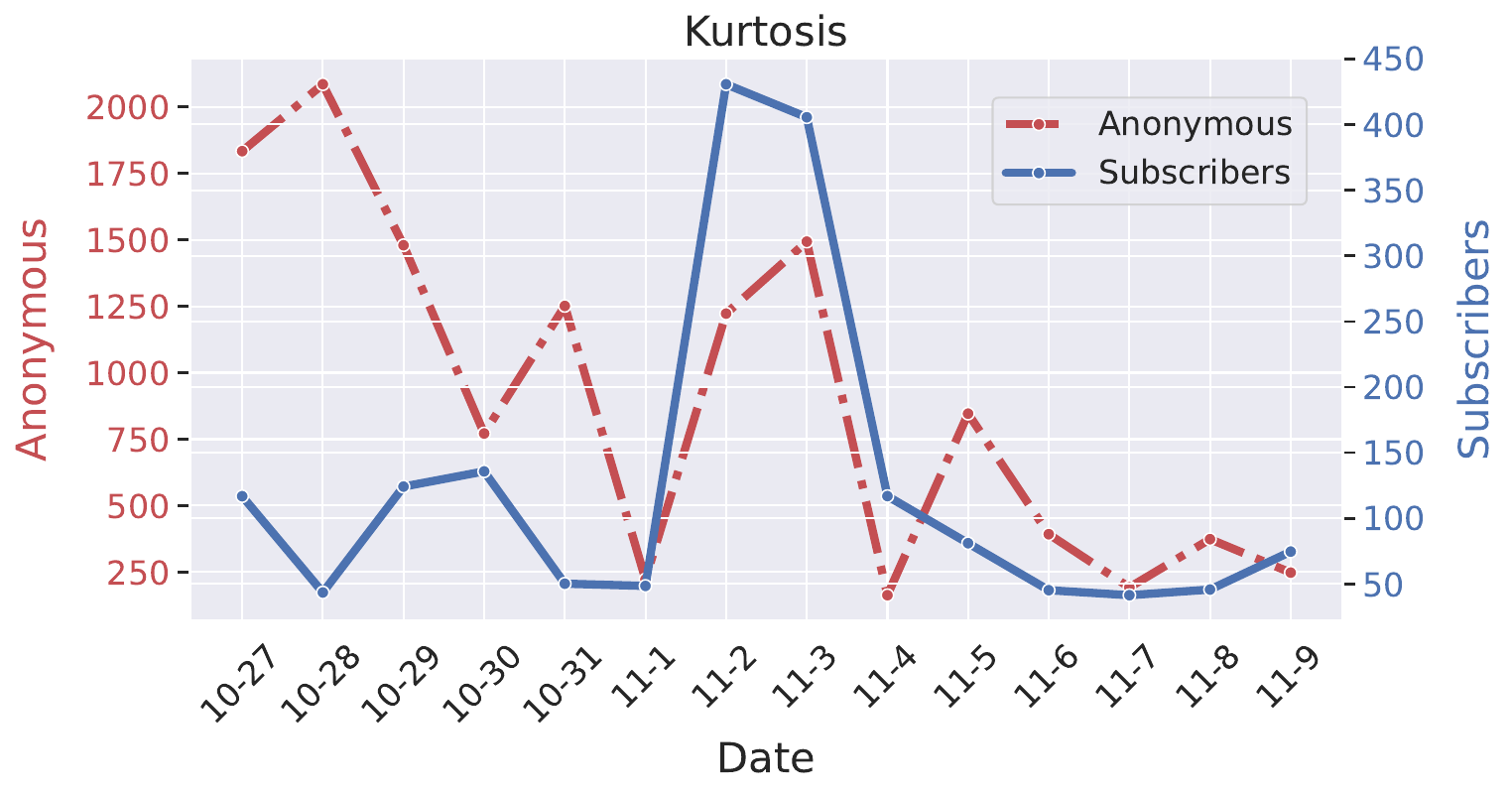}}

\caption{RQ2: Impact of personalized, content-based recommendations on the popularity bias in news article consumption measured by (a) skewness and (b) kurtosis based on the number of article reads for each day.} 
\vspace{-5mm}

\label{fig:pop_bias}
\end{figure}

The plots in Figure~\ref{fig:pop_bias} show the results addressing RQ2. For both metrics, i.e., skewness and kurtosis, we see a large gap between anonymous users and subscribers at the beginning of the study (i.e., 27th of October 2020), where only most-popular recommendations were shown to the users. While anonymous users have mainly read popular articles, subscribers were also interested in unpopular articles. This makes sense since subscribed users typically visit news portals for consuming articles within their area of interest, which will also include articles from the long-tail, while anonymous users typically visit news portals for getting a quick overview of recent events, which will mainly include popular articles. Based on this, a most-popular recommendation approach does not impact subscribers as much as it impacts anonymous users.

However, when looking at the last day of the study (i.e., 9th of November 2020), there is a considerably lower difference between anonymous and subscribed users anymore. We also see that the values at the beginning and at the end of the study are nearly the same in case of subscribed users, which shows that these users are not prone to popularity bias, and thus also personalized recommendations do not affect their reading behavior in this respect.  
With respect to RQ2, we find that the introduction of personalized recommendations can help to mitigate popularity bias in case of anonymous users. Furthermore, we see two significant peaks in the distributions that are in line with the COVID-19 lockdown announcement in Austria and the Vienna terror attack. Hence, in case of significant events also subscribed users are prone to popularity bias.

\section{Conclusion}
In this paper, we discussed the introduction of personalized, content-based news recommendations on DiePresse, a popular Austrian news platform, focusing on two specific aspects: user interface type (RQ1), and popularity bias mitigation (RQ2). 
With respect to RQ1, we find that the probability of recommendations to be seen is the highest for desktop devices, while the probability of clicking the recommendations is the highest for mobile devices.
With respect to RQ2, we find that personalized, content-based news recommendations result in a more balanced distribution of news articles' readership popularity for anonymous users. 
For future work, we plan to conduct a longer study, in which we also want to study the impact of different recommendation algorithms (e.g., use BERT~\cite{devlin2018bert} instead of LDA and include collaborative filtering) on converting anonymous users into paying subscribers. Furthermore, we plan to investigate other evaluation metrics, such as recommendation diversity, serendipity and novelty.

\vspace{1mm}
\noindent
\textbf{Acknowledgements.} This work was funded by the H2020 projects TRUSTS (GA: 871481), TRIPLE (GA: 863420), and the FFG COMET program. The authors want to thank Aliz Budapest for supporting the study execution.

\bibliographystyle{splncs04}
\bibliography{refs}

\end{document}